\documentclass{article}
\usepackage{spconf,amsmath,epsfig}
\usepackage{enumitem}
\usepackage{subfigure}
\usepackage{makecell}                 
\usepackage{booktabs}                 
\usepackage{tabu}                     
\usepackage{multicol}
\usepackage{multirow}
\usepackage{graphicx}
\usepackage{array}
\usepackage[psamsfonts]{amssymb}

\usepackage{threeparttable}
\usepackage{hyperref}
\usepackage{makecell}    
\usepackage{bbding}
\usepackage{pifont}
\usepackage{soul}

\hypersetup{
    colorlinks=true,
    linkcolor=black, 
    urlcolor=magenta 
}

\let\OLDthebibliography\thebibliography
\renewcommand\thebibliography[1]{
  \OLDthebibliography{#1}
  \setlength{\parskip}{0pt}
  \setlength{\itemsep}{0pt plus 0.3ex}
}

\begin{document}

\title{Multitask Frame-Level Learning for Few-Shot Sound Event Detection}
%
\name{
      Liang Zou$^{1}$,
      Genwei Yan$^{1,*}$,
      Ruoyu Wang$^{2}$,
      Jun Du$^{2}$,
      Meng Lei$^{1}$,
      Tian Gao$^{3}$, 
      Xin Fang$^{3}$
      }
\address{   
        $^1$ China University of Mining and Technology, Xuzhou, China, \{liangzou, gwyan, lmsiee\}@cumt.edu.cn\\ 
        $^2$ University of Science and Technology of China, Hefei, China, \\
        wangruoyu@mail.ustc.edu.cn, jundu@ustc.edu.cn\\ 
        $^3$ iFLYTEK Research Institute, Hefei, China, \{tiangao5, xinfang\}@iflytek.com\\ 
}
\maketitle

\begin{abstract}
This paper focuses on few-shot Sound Event Detection (SED), which aims to automatically recognize and classify sound events with limited samples. However, prevailing methods methods in few-shot SED predominantly rely on segment-level predictions, which often providing detailed, fine-grained predictions, particularly for events of brief duration. Although frame-level prediction strategies have been proposed to overcome these limitations, these strategies commonly face difficulties with prediction truncation caused by background noise. To alleviate this issue,  we introduces an innovative multitask frame-level SED framework. In addition,  we introduce TimeFilterAug, a linear timing mask for data augmentation, to increase the model's robustness and adaptability to diverse acoustic environments. The proposed method achieves a F-score of 63.8\%, securing the 1st rank in the few-shot bioacoustic event detection category of the Detection and Classification of Acoustic Scenes and Events Challenge 2023.
\end{abstract}

\section{Introduction}
\label{sec:intro}
Automatic Sound Event Detection (SED) involves identifying specific sound events in an audio clip and determining their start and end time. It has promising applications in various fields, such as smart home systems for security and automation~\cite{serizel2020sound}, environmental monitoring for detecting wildlife~\cite{you2023transformer} or urban sounds~\cite{9737390}, industrial settings for machinery monitoring~\cite{9054344}, and healthcare for patient monitoring and assistance~\cite{mesaros2019sound, mesaros2010acoustic}. Over the past decade, deep learning, particularly Convolutional Neural Networks (CNNs), has significantly advanced SED, yet implementing supervised CNN for SED requires extensive annotated data for each acoustic event category, which is both time-consuming and costly~\cite{wang2020few}. This is particularly problematic in specific applications like industrial machine condition monitoring, where collecting a broad spectrum of malfunction sounds is difficult. To address these challenges, the current research focuses on Few-Shot Sound Event Detection (FSSED), which aims to effectively classify sound events with a minimal number of samples, offering a more practical solution for applications where extensive data collection is not feasible~\cite{9054708}.

Different from conventional SED, in the FSSED setting, the model first undergoes initial pretraining on labeled data with base classes. Then, model generalization is evaluated on few-shot tasks, composed of unlabeled samples from novel classes unseen during training. In recent years, the FSSED task has attracted significant attention, and its methodologies fall into two categories. One relies on template matching~\cite{vanderbrug1977two}, which utilizes normalized similarity for event detection between training and query examples. For instance, Nolasco et al.~\cite{nolasco2023fewshot} employed a N-way K-shot approach to pretrained a meta lightweight feature extractor, where N is the number of classes and K is the number of known examples for each class. Then they performed event detection based on the normalization embedding features cross-correlation between the support and query segments. Another approach is based on prototypical networks (Protonet)~\cite{snell2017prototypical}, emphasizing rapid adaptation to novel classes using a few support examples. For instance, Tang et al. applied Protonet for extracting embedding features of sound events and incorporated an attention mechanism to focus on transient events and low-energy audio segments~\cite{tang2021two}. Mark et al. augmented data using time-frequency masks and time-domain stretching, enhancing the Protonet's generalization through contrastive learning~\cite{anderson2021bioacoustic}. Yang et al. employed transductive learning~\cite{boudiaf2020information} and Protonet to explore the deep relations between support and query set, and proposed a two-stage mutual-learning framework to further improve the network~\cite{yang2022mutual}. 

Despite the promise shown by existing methods in FSSED, they mainly depend on fixed segment-level prediction, which falls short in capturing short-term sound events. To address this concern, we introduced an adaptive frame-level detection method in the Detection and Classification of Acoustic Scenes and Events (DCASE) Challenge 2022, securing the 1st rank~\cite{zhang2023frame}. However, this method is   is susceptible to noise interference, often leading to the separation of single noisy events into two or more separate occurrences.

In response, we propose a novel multitask frame-level network for FSSED, featuring an additional binary classification branch for Sound Foreground and Background Classification (SFBC). The shared representations from multitask setup are able to capture more nuanced features from the audio. The SFBC branch, supplemented with a self-attention mechanism, enhances the model's ability to capture temporal information effectively. To further improve our FSSED model, we introduce TimeFilterAug, a novel data augmentation technique using a random linear filter mask. Our contributions are summarized as follows:

\vspace{-2pt}
\begin{enumerate}[label=(\arabic*), itemsep=-4pt]
    \item We introduce a pioneering multitask frame-level network for FSSED, incorporating an SFBC branch to address frame-level SED truncation issues.
    \item We propose the TimeFilterAug linear filter mask to augment the support and further improve the FSSED model.
    \item The developed frame-level system secure the top rank in the DCASE 2023 Challenge. The source code is  available at~\url{https://github.com/usefulbbs/Dcase2023Task5}. 
\end{enumerate}


\section{DATASET AND METHODOLOGY}
\subsection{Dataset}
Our experiments conducted on the DCASE2023 task 5 dataset~\cite{nolasco2023fewshot} including development and evaluation sets $X_{test1}$. The development set consists of predefined training $X_{base}$ and validation sets $X_{test2}$, with no overlap in terms of sound events $X_{base} \cap X_{test} = \emptyset $. The $X_{base}$ encompasses approximately about 21 hours of audio recordings across 47 bioacoustic classes, each accompanied by multi-class temporal annotations. Due to the presence of numerous incorrectly annotated events in the WMW subfolder of $X_{base}$, we exclude it from our analysis, focusing on the remaining 14 hours of audio containing 20 sound event classes. The $X_{test1}$ and $X_{test2}$ are composed of 66 and 18 audio files, respectively. Each file represents an independent few-shot task, providing annotations only for the initial five events. In the few-shot settings, the support set is denoted as $S =\left \{N_1,P_1,…,N_5,P_5\right \}$, where $S$ signifies the support set, and $P_i$ and $N_i$ represent the target events ($POS$) and background sounds ($NEG$), respectively. The challenge lies in detecting subsequent sound events based on these $POS$ annotations.

Considering there is no overlap between $X_{base}$ and $X_{test}$, we propose two unique frameworks tailored for the training and fine-tuning stages of our approach, which we detail in the subsequent sections.

\subsection{Multitask training framework}
\begin{figure}[t] 
    \centering
    \includegraphics[scale=0.78,angle=0]{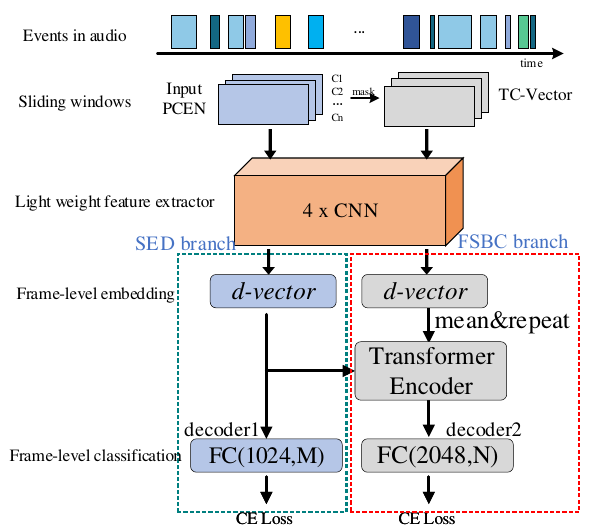}  
    \caption {Multitask frame-level embedding learning training framework. M is 20, N is 2. $C_n$ denotes the sequentially selected target class.}
    \label{trainFrameWork}
\end{figure}

Multitask learning effectively harnesses shared representations to discern commonalities across a spectrum of related tasks, a method proven to enhance data efficiency and mitigate inherent limitations of deep learning methodologies. In the training stage, we employ the multi-task training strategy to pretrain a lightweight feature extractor, as shown in Figure~\ref{trainFrameWork}. This setup includes two branches: the SED branch and the frame-level SFBC branch. In the SED branch, we first utilize Per-Channel Energy Normalisation (PCEN)~\cite{10095474} with a resampling rate of 22050hz, and segment the audios  into 5-second windows with 1-second overlaps. Then the PCEN windows are fed into the lightweight backbone, comprising 4 CNN\_Blocks (CNN+BN+ReLu), to produce SED embedding features. To get the frame-level prediction, we pad the embedding feature with repetitive values along the temporal axis. Furthermore, to avoid repetitive training of the overlap events between adjacent sliding windows, we train the overlapping segments between adjacent windows only once.

In the SFBC branch, we first sequentially select event class in the input window as the target class. Then we select a preceding window containing this class to avoid premature $POS$ location identification during the input samples feature extraction. After that, we mask the selected window based on the $POS$ frames location to construct the Target Class Vector (TC-Vector). The masked TC-Vector can be depicted in Eq~\ref{ts-vector}.


\begin{figure}[htbp] 
    \centering
    \includegraphics[scale=0.78,angle=0]{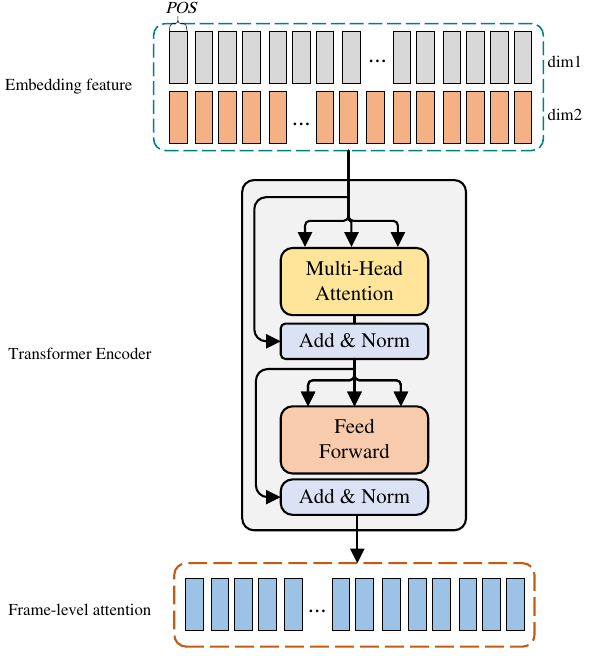}  
    \caption {The feature interaction in the Transformer Encoder. The repetitive $POS$ embeddings and SED embeddings are in the dim1 and dim2 channel, respectively.}
    \label{tfencoder}
\end{figure}

\begin{equation}
x^{k}_i = \begin{cases}
  x^{k}_i & \text{if } y_i = c_t \\
  0 & \text{else}
\end{cases}
\label{ts-vector}
\end{equation}
The $x^{k}_i$ denotes a frame of the input PCEN features, $k$ denotes 128 dimension, $y_i$ is its corresponding label, $c_t$ is the target class which is literally selected. To facilitate interaction between the two branches and capture temporal relationships in sounds, we stack the SED embedding and SFBC embedding along the channel dimension and feed them into one Transformer Encoder Layer, shown in Fig.~\ref{tfencoder}. Finally, we adopt CE loss in both SED branch and SFBC branch, denoted as $l_1$ and $l_2$, respectively. The $l_{total}$ denotes the total loss to minimize. They are calculated using the following equations:

\begin{equation}
    l_1=-\frac{1}{N} \sum_{i=1}^{n}\sum_{j=1}^{c}((Y_{i}==j)\odot {\mathbb{M}_i})\log(f_\phi({X_i}))
\label{loss}
\end{equation}

\begin{equation}
    l_2=-\frac{1}{N} \sum_{i=1}^{n}\sum_{j=0}^{1}((A_{i}==j)\odot {\mathbb{M}_i})\log(f_\phi({X_i},{X_t}))
\label{loss}
\end{equation}

\begin{equation}
    l_{total}=l_1 + l_2
\label{loss}
\end{equation}
$X_i$ represents the input PCEN window features, $Y_i=(y_1,y_2,...,y_n)$ is the corresponding frame-level label, $y_i$ $\in \left \{0,...,19 \right \}$, ``1$\sim$19" represents the label of 19 types of events in the training set, and ``0" represents background events. $c$ is the total number of classes (i.e., 20). $A_i=(a_1,a_2,...,a_i)$ denotes the frame-level SFBC label of the input window, ${a_i\in \left \{0,1 \right \}}$. ${\mathbb{M}_i} \in\left \{0,1  \right \}$ denotes the training mask for frames within a window, and ``0'' is for the overlapped frame of the same event between adjacent windows. '$\odot$' is the dot product function. $X_t$ represents the TC-Vector. ${N}$ is the total number of sliding windows.

\begin{figure}[t]
    \centering
        \subfigure[The SED branch]{
        \centering
        \label{SED_branch}
        \includegraphics[width=0.4\textwidth]{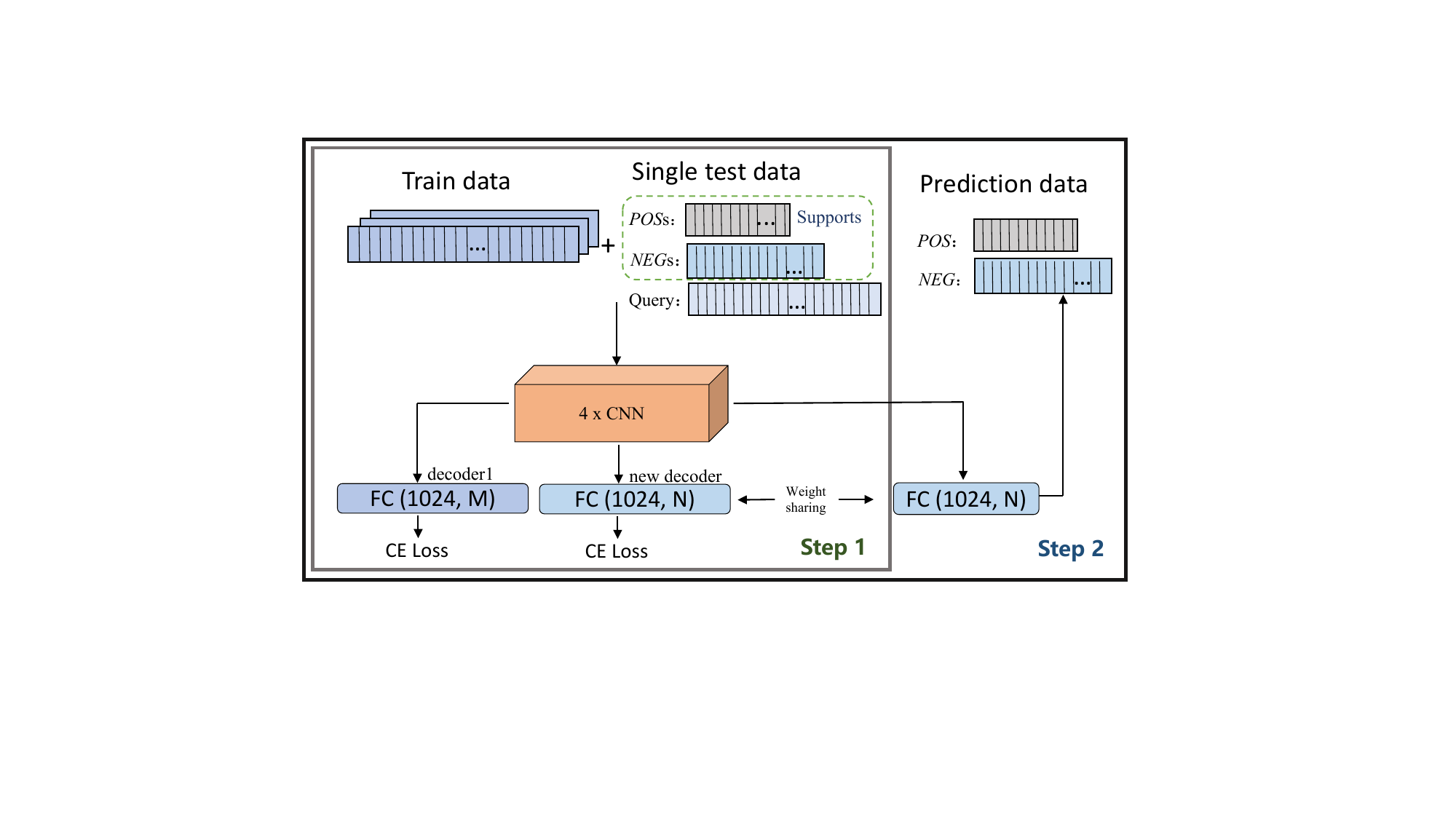}}

    \subfigure[The SFBC branch]{
        \label{TS_VAD_branch}
        \includegraphics[width=0.4\textwidth]{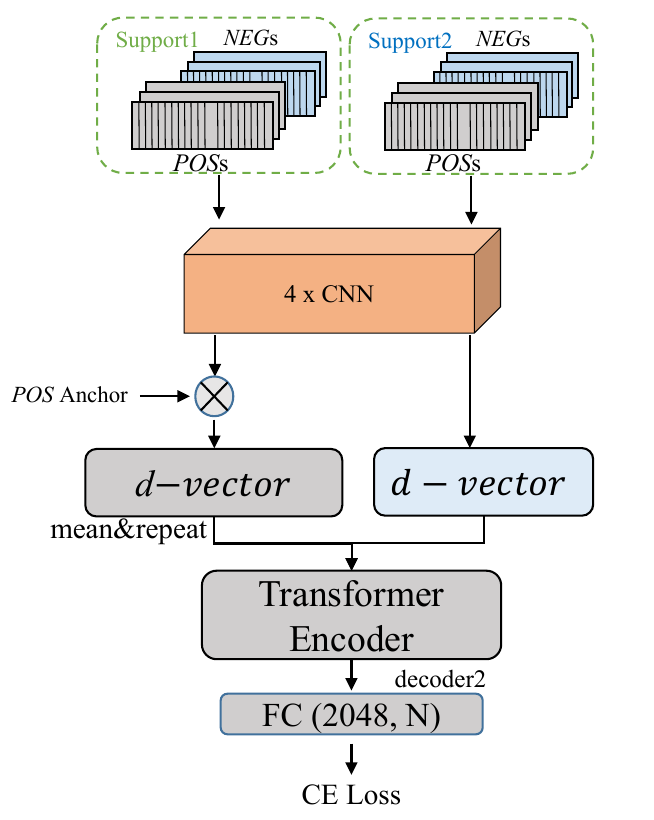}}
    \caption{Multitask frame-level learning fine-tuning framework.
    The $POS$ Anchor in (b) denotes the frames belonging to the sound event. M is 20, N is 2.}
    \label{fineTuneFrameWork}
    \end{figure}

\subsection{Multitask fine-tuning framework}
In the fine-tuning stage, we first enhance the diversity of our labeled training data through a process of support set reconstruction. Then, we fine-tune the SED branch and the SFBC branch respectively on the reconstructed supports. 

~\textbf{Supports reconstruction}
It is reasonable to hypothesize that in a frame-level system, sound event detection isn't highly reliant on the non-event class ($NEG$). We aim to enhance sample diversity by randomly altering the positions of $NEG$ and $POS$. Through random sampling and merging, we reconstruct the input samples in $Support_2$ and $Support_1$. Specifically, given a support set $S={\left \{N_1,P_1,...,N_5,P_5\right \}}$ and a query set $Q={\left \{q_1, q_2,...,q_i\right \}}$, where $N_i,P_i$ denotes the $NEG$ and $POS$ separately, and $q_i$ denotes the 5s sliding window on the query set, we feed each $q_i$ into the pre-trained feature extractor $f_\phi$. The cosine similarity of each $q_i$ to the $POS$ class is calculated as follows:

\begin{equation}
    \left \langle Z_q| Z_p  \right \rangle = Z_q\otimes \frac{1}{n} \sum_{i=1}^{n}Z_{P_i}
    \label{simliarity}
\end{equation}
$Z_q$ represents features of the sliding window in $Q$. $\frac{1}{n} \sum_{i=1}^{n}Z_{P_i}$ represents the class center of $POS$ features. '$\otimes$' represents the matrix multiplication. Subsequently, we enhance $N_i$ by selecting the low similarity $q_i$ to augment $N_i$. We further construct $Support_1$ and $Support_2$ (to mirror reality, only a single $POS$ is included in each sample window) by randomly sampling and concatenation from $P_i$ and the augmented $N_i$. 

~\textbf{Multitask fine-tuning} Similar to the training stage, in this stage, we utilize two multitask branches for the model fine-tuning, with distinct strategies for each branch. The fine-tuning framework is shown in Figure~\ref{fineTuneFrameWork}.

The SED branch adopts the two-step fine-tuning strategy. Initially, we define a  binary classifier based on the reconstructed supports set ($support_1$ and $support_2$). To enhance regularization, the training set is mixed with the $POS$ in supports set for a 20-classification task. The weight of 0-class in $decoder_1$ is replaced by the normalized embedding features of $POS$. Subsequently, we employ the pseudo-labels generated in the initial stage to further refine the model. This two-step process is iterated for a predetermined number of cycles.

In the SFBC branch, we select the windows in $Support_1$ containing multiple different $POS$ as TC-Vector and feed them into the lightweight feature extractor. A 0-1 masking based on the position of $POS$ frames in the window is applied. We obtain the $POS$ center by meaning the masked $POS$ frames and utilize the Transformer Encoder Layer transferred from the training stage to build a self-attention with the $POS$ in $Support_2$. A linear layer is followed to classify the frames in $Support_2$ based on the self-attention values. The pipeline is shown in Figure~\ref{TS_VAD_branch}. 

\subsection{TimeFilterAug}
Upon examining the experimental audios, we observe that the provided 5 shots are typically clear, whereas the query set for predictions often contains interference noise, predominantly from far-field sound and background impulse noise. To simulate these noisy conditions, we introduce TimeFilterAug, a random linear enhancement filter to simulate the noisy interferences. Specifically, we first randomly partition the 5-second PCEN sliding window into $m$ (ranges 3 to 6 in our work) segments, denoted as $T={\left \{T_1,T_2,...,T_m\right \}}$. Then, a random gain factor $g_i$ ranging from 0 to 1 is assigned for each $T_i$. Furthermore, we apply a linear mask varying between $-6dB\sim8dB$ for every $T_i$ based on $g_i$. The linear mask function is depicted as Eq~\ref{TimeFilterAug}. 
\begin{equation}\label{TimeFilterAug}
\begin{gathered}
\text{
 Aug\_}T_i = T_i \cdot 10^{(b/20)} \\
  \beta = l + (r-l) \cdot \alpha \\
  \alpha = \left[g_{i-1}, \ldots, g_i\right]
\end{gathered}
\end{equation}
'$\cdot$' is the element-wise multiplication. $\beta$ denotes dB gain factors ranging from $l$ to $r$. The parameter $l$ is the lower bound of gain, typically expressed as a negative value, while $r$ is the upper bound of gain, generally specified as a positive value. In this, we empirically set $l$ and $r$ to be -6 and 8, respectively. $\alpha$ is the linear space between two adjacent factors $g_i$. 
\section{EXPERIMENT}
\label{sec:typestyle}
\subsection{Experimental setups}
\noindent
\noindent
\textbf{Preprocessing} We extract PCEN features from the Mel spectrum at a 22050 Hz sampling rate, using 1024 n\_fft and 256 hop\_len. The sliding window segmentation of PCEN features is configured with a window length of 431 and a shift of 86, corresponding to 5 seconds and 1 second, respectively. Each frame within this window receives labeling based on the provided annotation. we employ Kaldi tools to manipulate the speed of the training audio. Specifically, we adjust the audio speed to 0.9x, 1.0x, and 1.1x its original rate. This modification alters both the timbre and intonation of the audio, providing a more varied and comprehensive dataset for training purposes. 

\noindent
\textbf{Training} We employ a 20-class and a 2-class classification on the training set, and optimize them using Adam optimizer with a learning rate of 1e-4, and employ StepLR with gamma 0.5 and step\_size 10. CE loss is adopted, excluding the overlap events labeled as 0. The number of iterations is set to be 100. The lightweight feature extractor consists of 4 CNN\_Blocks with 3$\times$ 3 kernel size and 128 input/output size. The Transformer Encoder includes 8 multi-head attention heads and 2 linear feed forward layers with 2048 input/output size.

\noindent
\textbf{Fine-tuning} We introduce a novel binary classifier based on the 0-1 ratio of the support set. The two decoders ($decoder_1$ and $decoder_2$) from the training phase are utilized. The class ``0'' weights in $decoder_1$ are replaced with the novel 2-classifier's class ``1” weights to refine the lightweight feature extractor. The fine-tuning stage involves training three decoders and the  last two layers of the encoder. We use Adam with learning rates of 1e-3 for SED and 1e-4 for SFBC, spanning 100 iterations. The TimefilterAug is applied after the 40th iteration, splitting the window into 6 time zones (i.e., $T_1{\sim}T_6$), with a minimum zone size of 48 frames. 

\begin{table}[tp]
\caption{Comparison of F-scores across various techniques on the DCASE 2023 Task5 Development and Evaluation dataset. FL and MFL denotes the single-task and multitask frame-level embedding learning, respectively.}
\centering
\renewcommand{\arraystretch}{1}
\tabcolsep=0.5cm
\begin{tabular}{c|c|c}\hline
Method &Eval-set  &Dev-set \\\hline
Wilkinghoff $et\,al.$~\cite{Wilkinghoff2023}  &16.0   &62.6\\\hline
Lee $et\,al.$~\cite{Jung2023}                 &27.1   &\textbf{81.5}\\\hline
Gelderblom $et\,al.$~\cite{Gelderblom2023}   &31.1   &36.6\\\hline
Liu $et\,al.$~\cite{XuQianHu2023}          &42.5   &63.9\\\hline
Moummad $et\,al.$ ~\cite{Moummad2023}     &42.7   &63.5\\\hline
FL~\cite{zhang2023frame}     & -   &70.2\\\hline
\textbf{MFL(ours)}      &\textbf{63.8}   &77.25\\\hline
\end{tabular}
\label{Fscore_final}
\end{table}
\begin{table}[tbp]
\caption{The ablation study on the modification methods of single task frame-level system. BS represents the balance sampling.}
\centering
\renewcommand{\arraystretch}{1}
\tabcolsep=0.45cm
\setlength{\tabcolsep}{1.1mm}{
\begin{tabular}{c|ccc}\hline
 \multicolumn{1}{c|}{Modification}   & \multicolumn{1}{c}{Precision}  &\multicolumn{1}{c}{Recall}  &\multicolumn{1}{c}{F-score}  \\ \hline
    -                     &$75.73(\pm2.15)$                  &$63.1(\pm2.71)$     &$68.8(\pm1.67)$   \\ 
    BS                    &$76.20(\pm0.86)$                  &$65.15(\pm0.79)$    &$70.67(\pm0.80)$   \\ 
    KFold                 &$74.67(\pm2.93)$                  &$67.91(\pm3.0)$     &$70.37(\pm2.03)$   \\ 
    KFold+BS              &\textbf{76.26($\pm$0.64)}   &\textbf{68.51($\pm$0.66)}    &\textbf{72.17($\pm$0.37)}   \\ \hline
\end{tabular}}
\label{modification}
\end{table}
\subsection{Experimental results}
Table~\ref{Fscore_final} presents results from the DCASE 2023 Task5. Our innovative frame-level system attains a notable F-score of 63.8\% on the evaluation dataset, surpassing competing methods~\cite{Moummad2023}. The prevailing methods focus on segment-level predictions, limiting their efficiency in handling short-time events. In contrast, the proposed multi-task frame-level network achieves more accurate results through enhanced adaptive frame-level prediction. In addition, the multitask framework shows a substantial improvement of 7.05\% over the single-task framework on the development set., as shown in the last two lines of Table~\ref{Fscore_final}.

\subsection{Modification on frame-level system}
Instabilities were identified in the frame-level FSSED system, mainly due to data imbalances, particularly in brief sounds, as shown in the first line of Table~\ref{modification}. This instability arises from the data imbalance within the sample window during the fine-tuning phase, notably for the brief sounds in PB subfolder of validation set. For instance, the aggregate duration of 5-shot for such sounds is merely 50ms, while the background noise spans 20s. This imbalance can misdirect the network, causing it to concentrate on background noise while overlooking the target sound. To counter this, we utilized the random over-sampling technique to guarantee an equal duration of the target sound in every sample. To maximize the effectiveness of our limited dataset, we integrated KFold cross-validation into our model training regimen. We evaluate and select the best-performing model from each fold. Subsequently, in the testing phase, we fuse the results from these individually optimized models. The results in Table~\ref{modification} show the effectiveness of these strategies in stabilizing the results. 

\begin{table}[t]
\begin{center}
\caption{Ablation study on the effect of different methods on frame-level embedding learning framework. Seq-model denotes the sequential model.}
\begin{tabular}{c|ccc|c}
\Xhline{1pt}
\multirow{2}{*}{} & \multicolumn{3}{c|}{Methods}      & \multicolumn{1}{c}{Dev-set} \\ \cline{2-5}
        & Augmentation        &Seq-model        &Task              & F-score         \\ \hline
 1      & -          & -                &SingleTask                        & 72.17     \\ 
 2      & -          & -                &Multitask                         & 73.14     \\ 
 3      & -          & LSTM              &Multitask                         & 74.43     \\ 
 4      & -          & Transformer       &Multitask                         & 76.40     \\ 
 5      & TimeFilterAug        & Transformer       &Multitask                         & \textbf{77.25}  \\ 
\Xhline{1pt}
\end{tabular}
\label{results}
\end{center}
\end{table}

\subsection{Ablation study}
\label{sec:typestyle}
We evaluate the influence of various modifications on the frame-level embedding learning framework on the development set.

\textbf{Influence of Multitask Learning} Introducing multitask learning with the addition of the SFBC branch (without sequential modules) improves the model's performance from 72.17\% to 73.14\%, as shown in line 1 and line 2 of Table~\ref{results}. 

\textbf{Influence of Transformer Encoder} Incorporating the LSTM module at the end of the feature extractor is able to assist the model in capturing context and facilitate interactions between different tasks, especially in the SFBC branch where temporal context is crucial. The F-score is improved by 1.29\% (as evidenced by line 3 in Table~\ref{results}) over 73.14\% validates this assertion. Furthermore, replacing the LSTM with a transformer encoding layer further improve the performance, leading to a significant F-score of 76.40\%. This increase underscores the Transformer's superior ability to discern and process the temporal dynamics inherent in the data.

\textbf{Influence of TimeFilterAug} The introduction of TimeFilterAug results in a notable increase of 0.85\% in the F-score on the development set, as evidenced by the comparison between line 4 and line 5 in Table~\ref{results}. Remarkably, its impact is even more pronounced on the validation set, where it yields an enhancement ranging between 3\% to 5\%. This significant improvement underscores TimeFilterAug's capacity for effective generalization across diverse datasets. It should be mentioned that, due to the validation set not being publicly accessible, we are unable to perform ablation experiments to further investigate this aspect.

\section{CONCLUSIONS}
\label{sec:typestyle}
In this study, we introduce an advanced frame-level embedding learning framework which strategically leverages multitask learning to tackle the challenge inherent in existing FSSED methods. The systematic experiments have clearly demonstrated the superiority of this multitask methodology over the conventional single-task approaches. In addition, TimeFilterAug provides impressive performance boosts, especially in noisy scenarios. This innovative strategy has earned us a prestigious first-place finish in the Task 5 of DCASE 2023 Challenge, achieving an outstanding F-score of 63.8\% in the hold-out evaluation dataset.  Looking to the future, we are committed to further refining our system by exploring a range of innovative techniques to to further advance our capabilities in sound event detection.


\end{document}